\documentclass[sigconf]{acmart}

\usepackage{enumitem}
\usepackage{multirow}
\usepackage{graphicx}
\usepackage{balance}
\usepackage[caption=false]{subfig}
\usepackage{xcolor}
\definecolor{boxgrey}{HTML}{F3F3F3}

\AtBeginDocument{%
  \providecommand\BibTeX{{%
    \normalfont B\kern-0.5em{\scshape i\kern-0.25em b}\kern-0.8em\TeX}}}

\copyrightyear{2022}
\acmYear{2022}
\setcopyright{acmcopyright}
\acmConference[L@S '22] {Proceedings of the Ninth ACM Conference on Learning @ Scale}{June 1--3, 2022}{New York City, NY, USA.}
\acmBooktitle{Proceedings of the Ninth ACM Conference on Learning @ Scale (L@S '22), June 1--3, 2022, New York City, NY, USA}
\acmPrice{15.00}
\acmISBN{978-1-4503-9158-0/22/06}
\acmDOI{10.1145/3491140.3528273}

\begin{document}
\fancyhead{}

\title{Meta Transfer Learning for Early Success Prediction in MOOCs}

\author{Vinitra Swamy}
\affiliation{EPFL}
\email{vinitra.swamy@epfl.ch}

\author{Mirko Marras}
\affiliation{University of Cagliari}
\email{mirko.marras@acm.org}

\author{Tanja Käser}
\affiliation{EPFL}
\email{tanja.kaeser@epfl.ch}

\begin{abstract}
Despite the increasing popularity of massive open online courses (MOOCs), many suffer from high dropout and low success rates. Early prediction of student success for targeted intervention is therefore essential to ensure no student is left behind in a course. There exists a large body of research in success prediction for MOOCs, focusing mainly on training models from scratch for individual courses. This setting is impractical in early success prediction as the performance of a student is only known at the end of the course. In this paper, we aim to create early success prediction models that can be transferred between MOOCs from different domains and topics. To do so, we present three novel strategies for transfer: 1) pre-training a model on a large set of diverse courses, 2) leveraging the pre-trained model by including meta information about courses, and 3) fine-tuning the model on previous course iterations. Our experiments on $26$ MOOCs with over $145{,}000$ combined enrollments and millions of interactions show that models combining interaction data and course information have comparable or better performance than models which have access to previous iterations of the course. With these models, we aim to effectively enable educators to warm-start their predictions for new and ongoing courses. 
\end{abstract}



\maketitle

\section{Introduction}
Massive Open Online Courses (MOOCs) and digital learning environments are becoming immensely popular \cite{kruchinin2019investigation, ByTheNum39}. Enrollments have surged to over $220$ million MOOC learners, with several courses on Coursera and EdX reporting increases by a factor of $10$ due to the COVID pandemic \cite{impey2021moocs, ByTheNum39}. In 2021 alone, $40$ million new learners signed up for a MOOC, with providers hosting $3{,}100$ new courses and over $500$ new programs \cite{ByTheNum39}. However, MOOCs tend to suffer from high dropout and low success rates \cite{aldowah2020factors,MOOCcomp77}. Adaptive guidance can hence increase retention and student performance. Methods for \textit{early} success prediction build the basis for interventions \cite{perez2021can}.

Consequently, a large body of work has focused on student success prediction in MOOCs based on students' interaction data, using logistic regression \cite{whitehill2017mooc}, support vector machines \cite{tomkins2016predicting}, random forests \cite{sweeney2016next}, or long short-term memory (LSTM) networks \cite{wang2017deep}. Prior work has also suggested feature sets for solving the success prediction task. For example, \cite{chen2020utilizing, lemay2020grade} extracted count-based features (e.g. number of online sessions, number of videos watched). Others extracted features related to the course attendance rate or the ratio of videos watched \cite{he18utilize, mbo20early} and fine-grained video behavior (either through explicit features or through feeding raw video events into an LSTM \cite{mu21deeplearning}). \cite{boroujeni2016quantify} has showed that regular working patterns are essential for success in MOOCs. Notably, most of these studies have performed a-posteriori or post-hoc analyses. Only few works have addressed \textit{early} success prediction for ongoing courses. For example, \cite{marras2021can} performed a meta-analysis of features across early prediction models on flipped courses and MOOCs. \cite{mbo20early} showed that video engagement in the first week and mid-point of the course can be indicative of student pass-fail grades. \cite{mao2019one} used recent temporal patterns in novice programming tasks to early intervene.

The vast majority of performance prediction literature has focused on inference on the course the model is trained on, i.e. on training the models on one portion of the students and then predicting on the remaining students of the course. In case of first time smaller courses, training on the same course might lead to overfitting. For \textit{early} success prediction, using the same course for training and inference is not practical as the model does not have the data required for supervised training (the pass-fail label will only be known at the end of the course). Thus, for these settings, success predictions need to rely on models trained on other courses. 

 \textit{Transfer learning} is a well-known machine learning paradigm, which refers to training a model on one setting, and using that model to warm-start predictions in a different context \cite{weiss2016survey}. In the context of education, only a few works have attempted to create generalizable models that can be transferred between different settings. \cite{ding2019transfer} extract latent feature representations that are transferable over multiple MOOC courses. \cite{wei2017convolution} and \cite{karumbaiah2021using} warm-start their model on past data in two different settings: MOOC forum post classification and active learning tasks. \cite{tsiakmaki2020transfer} transferred predictions across 5 undergraduate courses at the same university. Most relevantly, \cite{whitehill2017delving} showed that learning generalized patterns of student behavior from multiple MOOCs can improve predictive performance.
 
In this paper, we aim to tackle the problem of transferability across MOOCs from different domains and topics, focusing on models for early success prediction. We present and analyze three novel strategies to creating generalizable models for this task. First, we train a BiLSTM model taking students' interaction data as input on a large collection of diverse MOOCs. Our second strategy is based on the assumption that students' interaction behavior is influenced by the context and hence courses with similar characteristics will have similar student interaction patterns. We suggest two architectures for including course information into a model, one being an extension of our previous BiLSTM model and the second one combining interaction data and course information using attention layers. We frame success prediction as a \textit{meta learning} problem (see e.g. \cite{hospedales2020meta}) through viewing broad domains of MOOCs as separate tasks i.e. predicting student success in a Mathematics course is often very different from predicting student success in an English course. Our third strategy extends the aforementioned models by applying fine-tuning on previous course iterations.

\begin{figure*}[t]
  \includegraphics[width=\textwidth, trim=1 1 1 1,clip]{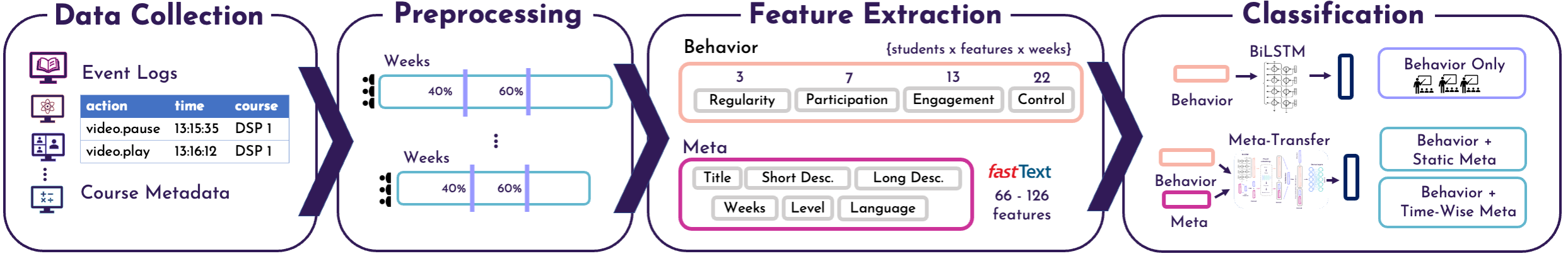}
  \vspace{-6mm}
  \caption{Our transfer approach based on behavior and course metadata features, from logs collection to success prediction.}
  \label{fig:features}
\end{figure*}

We extensively evaluate our models on a large data set including $26$ MOOCs and $145{,}714$ students in total. With our analyses, we address the following three research questions: 1) Can student behavior transfer across iterations of the same course and across different courses? 2) Is a meta learning model trained on a combination of behavior and course metadata information more transferable? 3) Can fine-tuning a combined model on past iterations of an unseen course lead to better transferable models? Our results demonstrate that a model statically combining behavior and meta information over multiple courses can significantly improve over one-course transfer baselines ($11.45\%$ on average across multiple early prediction levels and courses). Our implementation and pre-trained models\footnote{\texttt{https://github.com/epfl-ml4ed/meta-transfer-learning}} achieve at least a comparable and often better performance to training a customized model based on previous iterations of their course.

\section{Methodology}
The problem we address in this paper can be formulated as a time-series classification task that relies on students' interaction data as well as meta information about the course to predict \textit{early} on whether a student will pass or fail a course. Specifically, we are interested in improving the accuracy of success predictors for students belonging to a previously unseen iteration of a course or to a completely new course. For clarity, we formalize the addressed problem before presenting the proposed approach.

\subsection{Problem Formulation}
We consider a set of \emph{students} $S^c \subset \mathbb{S}$, who are enrolled in a \emph{course} $c$ (referred also as course iteration), which is part of the online educational offering $\mathbb{C}$. As courses can be run multiple times over the year, each course $c$ refers to one run of a \emph{course set} $C$. We denote a sequence of iterations for a course set $C = \{c_1, \ldots, c_{M^C}\}$, with $M^C$ being the total number of iterations for the course set $C  \in \mathbb{C}$. In other words, we consider a course set $C$ to be a set composed of subsequent course iterations (e.g., multiple runs of a linear algebra course). 
Each course $c \in C$ has a predefined \emph{schedule} consisting of $N_c = |\mathbb{O}^c|$ learning objects from a catalog $\mathbb{O}^c$. In our study, we assume that the latter can be either videos or quizzes, but the notation can be easily extend to other types (e.g., forum posts, textbook readings). 
Enrolled students interact with the learning objects in the schedule, generating a time-wise clickstream. We denote the \emph{interactions} of a student $s \in \mathbb{S}^c$ in a course $c$ as a time series $I_s^c = \{i_1, \ldots, i_K\}$ (e.g., a sequence of video plays and pauses, quiz submissions), with $K$ being the total number of interactions of student $s$ in course $c$. 
We leave these definitions very general, allowing the length of each time series to differ, since our models can accommodate this. We assume that each interaction $i$ is represented by a tuple $(t, a, o)$, including a \emph{timestamp} $t$, an \emph{action} $a$ (videos: load, play, pause, stop, seek, speed; quiz: submit), and a \emph{learning object} $o \in \mathbb{O}^c$ (video, quiz). 
We finally denote the \emph{binary success labels} (pass/fail) for students $S^c$ in a course $c$ as $\mathbb{L}^c = \{l_{s_1}, \ldots, l_{s_{|S^c|}}\}$.

\subsection{Transfer Framework for Success Prediction}
As shown in Fig. \ref{fig:features}, our approach to solving the transfer problem in early success prediction\footnote{Our approach and resulting models can be easily leveraged for other downstream educational machine learning tasks, e.g., grade prediction and dropout prediction.} consists of three steps: 1) collecting and preprocessing the log data, 2) computing behavior features for each student (based on log data) and meta features per course, and 3) building three different deep learning models based on different combinations of behavior and meta features.

\subsubsection{Data Collection and Preprocessing}
\label{sec:log-preproc}
A large portion of students drop out of a course during the first few weeks (e.g., there are students who just enroll in the course for watching a few videos) \cite{moocdropout, moocretention}. Therefore, it is possible to predict course success for these early-dropout students with a high accuracy by simply looking at their assignment (or homework/quiz) submissions and grades in the first course weeks. Developing complex machine learning models for predicting on these early-dropout students is inefficient. Furthermore, training a machine learning model on a highly imbalanced data set (i.e., with a large number of students failing the course) might lead to a biased model. For that reason, when not stated otherwise, we refer to $S^c$ as the students in a course $c$, whose course pass-fail label cannot be easily predicted with a shallow machine learning model, trained on the assignments grades in the first few weeks of the course. While we filter out so called early-dropout students, we will later show (Sec. \ref{sec:results-metatransfer}) that our predictors have high accuracy on these students as well. Our setting consequently leads to a faster and more effective optimization process. To identify the early-dropout students $s \in S^c$ of each course $c$, we fit a \textit{Logistic Regression} model on the assignment grades of the first course weeks. Specifically, our input data for the model is a matrix $G$ of shape $|S^c| \times w$, where $w$ is the number of course weeks and the value $G(i, j)$ is the grade student $s_i$ received on the assignment in course week $j$. In our case, $w = 2$. We considered the average score of graded assignments for that week in case of multiple graded assignments per week scored non-attempted assignments with a $0$. We filter out the students whose predicted probability of course failure $\hat{p}_{s}>0.99$. We found the optimal threshold via a grid search over $\{0.96, 0.97, \ldots, 0.999\}$, maximizing the model's \textit{balanced} accuracy.

Given the low success rates for many MOOCs, targeted interventions have the potential to improve learning outcomes \cite{borrella2021taking, xing2019dropout, whitehill2015beyond, perez2021can}. We are therefore interested in \textit{early} success prediction, providing the basis for such targeted intervention (e.g., offering additional support to students at risk of failing the course). We define an early prediction level $e$ and consider students' interaction data only up to that point in time \footnote{Our experiments will show early predictions for $40\%$ and $60\%$ of the course duration.}. If a course had a duration of $10$ weeks and we aimed to predict student success after $e = 40\%$ of the course duration, we would only consider student interactions happening in the first four course weeks. We denote the interactions of a student $s \in \mathbb{S}^c$ in a course $c$ up to a specific point in time $e$ with $I_s^{c,e}$.

\subsubsection{Feature Extraction}\label{sec:features-descr}
For each course iteration $c \in \mathbb{C}$, we extract a set of behavior features for each student $s$ based on the student's interaction data $I_s^{c,e}$ for an early prediction level $e$. We also extract a set of meta information about that course iteration $c$.

\vspace{1mm} \noindent \textbf{Behavior Features}. To obtain a comprehensive representation of student behavior, we combine multiple types of features according to student interactions with videos and quizzes. More granularly, we consider the four behavior feature sets identified as the ones with the highest predictive power for success prediction in the context of MOOCs \cite{marras2021can}\footnote{Given the size and variety of the course data considered in our study, we included all features of the four features sets, instead of considering only the specific features identified as important in at least one course by \cite{marras2021can}. We then let our classifiers detect the patterns and hence relevant features for the considered courses.}.  
Formally, given the interactions $I_s^{c,e}$ generated by students $S^c$ until a course week $w \in \mathbb{N}$ in course $c$, we produce a matrix $H \subset \mathbb{R}^{|S^c| \times w \times h}$ (i.e, each feature in the feature set is computed per student per week), where $h \in \mathbb{N}$ is the dimensionality of the feature set. As in Sec. \ref{sec:log-preproc}, $w$ is chosen based on the early prediction level $e$. We focus on the following behavioral aspects\footnote{These 45 features are explicitly detailed in Appendix \ref{sec:features}.}:

\begin{itemize}[leftmargin=*,nolistsep]
    \item \textbf{Regularity} \cite{boroujeni2016quantify} ($H_1$, shape: $|S^c| \times w \times 3$). These features quantify the regularity of a student's study habits, for instance by measuring whether the student usually studies during the same time of a day or has a preference for specific days of the week.
    \item \textbf{Engagement} \cite{chen2020utilizing} ($H_2$, shape: $|S^c| \times w \times 13$). These features monitor the engagement of students throughout the course, considering indicators such as the total number of student clicks on weekends and on weekdays, and the total number of sessions.
    \item \textbf{Control} \cite{lalle2020data} ($H_3$, shape: $|S^c| \times w \times 22$). This feature set measures the fine-grained video consumption per student, including features such as the proportion of videos watched, re-watched, or interrupted, and the standard deviation of these ratios.
    \item \textbf{Participation} \cite{marras2021can} ($H_4$, shape: $|S^c| \times w \times 7$). These features are related to attendance on videos and quizzes based on the schedule, such as the number of scheduled videos watched for that week and the number of quizzes passed on the first try.
\end{itemize}

\vspace{2mm} Given a course $c \in \mathbb{C}$, we extract the above features for all students $S^c \subset \mathbb{S}$ and concatenate them to obtain the final combined behavior features for students in that course, defined as $H^c \in \mathbb{R}^{|S^c| \times w \times 45}$, with $H^c = [H_1 \cdot H_2 \cdot H_3 \cdot H_4]$ (the $\cdot$ denotes a concatenation). Due to the different scales, we perform a min-max normalization per feature in $H^c$ (i.e., we scale the feature between 0 and 1 considering all students and weeks for that feature).

\vspace{1mm} \noindent \textbf{Meta Features}. Our application scenario includes a large variety of courses with different characteristics (e.g., topic, structure, duration). We assume that the observed interaction patterns of a course do not only depend on the population (i.e. the individual students attending the course), but also on the specific characteristics of the course. In prior work \cite{park2020meta, sun2019meta}, we see that passing in meta information representing differences in context improves predictive performance of transfer. We therefore code relevant information about each course. Formally, given a course $c \in \mathbb{C}$, we produce a fixed-length representation $F \subset \mathbb{R}^{f}$ for each meta feature, where $f \in \mathbb{N}$ is the size of the meta feature. Specifically, we consider:

\begin{itemize}[leftmargin=*,nolistsep]
    \item \textbf{Duration} ($F_1$, shape: $f=1$) represents the number of course weeks as per the schedule provided by the instructor.  
    \item \textbf{Level} ($F_2$, shape: $f=3$) is a one-hot encoded vector representing the level of the course. Our course data includes three different levels: Bachelor, Master, and Propedeutic.
    \item \textbf{Language} ($F_3$, shape: $f=2$) is a one-hot encoded vector representing the language in which the course is taught. Courses included in this study, were taught either in French and English. 
    \item \textbf{Title} ($F_4$, shape: $f=30$) is a word embedding vector of the title the instructor assigned to the course. 
    \item \textbf{Short Description} ($F_5$, shape: $f \in [30,60]$) is a word embedding vector of the short description of the course (one sentence about the main topic).
    \item \textbf{Long Description} ($F_6$, shape:  $f \in [30,60]$) is a word embedding vector of the long description of the course ($2{-}5$ more detailed sentences about the course content). 
\end{itemize}

\begin{figure*}[t]
  \includegraphics[width=\textwidth, trim=4 4 4 4,clip]{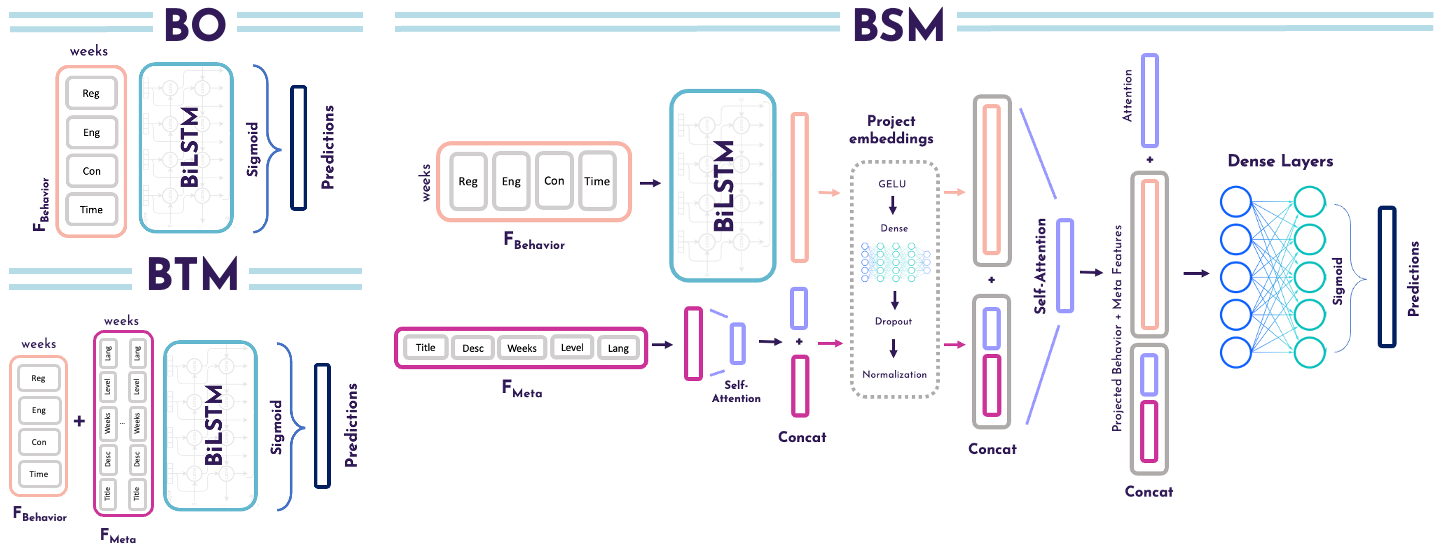}
  \vspace{-13mm}
  \caption{Model architectures adopted in our study.}
  \label{fig:models}
\end{figure*}

\vspace{2mm}The features representing the Title, the Short Description, and the Long Description were obtained by translating the corresponding text to English via the DeepL translation service and then fine-tuning a pre-trained FastText model \cite{fasttext} on that specific meta information\footnote{We experimented with the size of the latent vector space and the encoding frameworks and found that FastText outperformed other out-of-the-box encoders for this use case (SentenceBERT \cite{reimers2019sentencebert} and Universal Sentence Encoder \cite{cer2018universal}).}. FastText is a state-of-the-art model for word representation (i.e., representing words in a text as latent vectors that can be used for machine learning classification tasks). Intuitively, we expect the Duration, Language, and Level meta features to encode similarities in student population across courses and the Topic and Description meta features to reflect similarity on course content. Given a course $c \in \mathbb{C}$, we extract the above features and concatenate them to obtain the final combined meta features for that course. Formally, the meta features for course $c$ are denoted as $F^c \in \mathbb{R}^{124}$, with $F^c = [F_1 \cdot \ldots \cdot F_6]$. Again, due to the different scales of these meta features, the values in $F^c$ are min-max normalized per feature. 

\subsubsection{Classification}\label{sec:model-creation}
In our transfer setting, given a set of courses $C_{trn} \subset \mathbb{C}$, we are interested in creating an accurate success prediction model trained on $C_{trn}$ for a specific early prediction level $e$, which can be transferred to new (unseen) courses or new (unseen) iterations of a course. We are interested in creating models that can predict accurately on courses not seen during training, i.e., $C_{trs} \subset \mathbb{C}$, with $C_{trs} \cap C_{trn} = \emptyset$ and $C_{trs} \cup C_{trn} = \mathbb{C}$. For a given early prediction level $e$, we investigate the three model architectures shown in Fig. \ref{fig:models}. 

\vspace{2mm} \noindent \textbf{Behavior Only} (\texttt{BO}). Our first model predicts student success based on behavior features (and thus student interaction data) only. The architecture of this model is illustrated in Fig. \ref{fig:models}. Formally, the model input is defined as $H = \cup \; H^c\; \forall \;c \in C_{trn}$. Given that each course in $C_{trn}$ potentially has a different number of weeks, we consider the maximum course duration $\textbf{w} = \max(\cup  \; w^c) \; \forall \; c \in C_{trn}$ and pad behavior features for shorter courses with $-1$. The input $H$ therefore has a shape of $|S^c|$ $\times$ $\textbf{w}$ $\times$ $45$. There features are then fed into a neural architecture composed by a \emph{Masking} layer ($mask=-1$), a simple yet effective BiLSTM layer\footnote{We experimented with traditional machine learning models (e.g., Support Vector Machines, Logistic Regression, Random Forest) and deep-learning models (e.g., Dense Fully-Connected Networks, RNNs, LSTMs, CNNs, and BiLSTMs), and found that the \emph{BiLSTM} layer performs best against the other baselines for our use case.} with a loopback of $3$, and a \emph{Dense} layer (with Sigmoid activation) with a hidden size of $1$. The model outputs the probability the student will fail the course.

\vspace{1mm} \noindent \textbf{Behavior + Time-wise Meta} (\texttt{BTM}). Our second model is based on the assumption that students' interaction behavior is influenced by the specific characteristics of the course. We expect that courses with similar characteristics in terms of structure (e.g., topic, duration) and level (e.g., Bachelor) exhibit more similar behavior patterns than courses with completely different characteristics. Therefore, including meta features into the model might improve transfer (i.e. predictive performance of the model on unseen iterations or courses). The resulting model is illustrated in Fig. \ref{fig:models}. We use the exact same architecture as for the \texttt{BO} model; we only change the input data of the model to be a concatenation of both behavior and meta features for each week, formally denoted as $H = \cup \; H^c \cdot F^c \; \forall \; c \in C_{trn}$ (with $\cdot$ representing a concatenation operator). The meta features fed into the model are the same for every week since they are computed per course. The final input $H$ has a shape of $|S^c|$ $\times$ $\textbf{w}$ $\times$ $169$ ($45$ for the behavior features and $124$ for the meta features). By feeding in the meta features for every week, we allow the BiLSTM to directly learn from the dependencies between meta features and behavior features over time.

\vspace{1mm} \noindent \textbf{Behavior + Static Meta} (\texttt{BSM}). Similar to our second model, our third model also relies on the assumption that students' interaction behavior varies dependent on the course characteristics. However, this third model is based on the hypothesis that not all course features are equally important, i.e. some characteristics might have a much influence on student behavior than others. We hence add attention to this model. The architecture consists of separate branches for the behavior feature and meta features, followed by a head architecture that elaborates the combined behavior and meta features (Fig. \ref{fig:models}). The behavior features branch has the same architecture as the \texttt{BO} model, except for the removal of the final \emph{Dense} layer. 

The meta feature branch receives the meta features $F^c$ of the course $c \in C_{trn}$ (shape: $|S^c| \times |F^c|$) as input and is composed of two layers. The first layer is a Bahdanau \emph{Attention} layer \cite{bahdanau2014neural} which returns the weights for all features in $F^c$, representing the importance of each of these features for the prediction, i.e., a vector of size $|F^c|$). The second layer is a \emph{Concatenation} layer that combines both raw meta features and their importance weights returned by the \emph{Attention} layer (shape: $2 \times |F^c|$). This attentive layer is essential for internal and external model interpretability, making it possible to identify the most important (meta) features for the prediction.

Behavior and meta features follow different scales and variation patterns and are differently associated to each input instance (e.g., meta features are the same for each student belonging to a given course, while behavior features change for each student). Therefore, the head architecture combining both the behavior branch and the meta branch does not direct rely on a concatenation of the outputs of these two branches. Rather, the output of each branch is individually projected to the same latent space, obtaining a latent vector (shape: $256$) for the output of each branch. To this end, we use a well-known \emph{Projection} block, composed by a sequence of GELU, Dense, Dropout, and Normalization layers, as proposed by \cite{Multimod35:online}. The two latent vectors are then fed into a \emph{Concatenation} layer (shape: $2 \times 256$). A second Bahdanau \emph{Attention} layer returns a vector of size $2 \times 256 = 512$ representing the weights for all latent behavior and meta features. This second attentive layer weights the overall importance of behavior and of meta information for each prediction. Then, another \emph{Concatenation} layer combines both the latent features and their importance weights (shape: $2 \times 512 = 1024$).  Finally, the concatenated output is passed to a cascade of \emph{Dense} layers, with the last one having a Sigmoid activation and a hidden size of $1$. This final layer returns the probability the student will fail the course. 

\section{Experimental Evaluation}\label{sec:results}

We conducted experiments
to understand the extent to which student behavior transfers well on iterations of the same course or of a different course (RQ1), the extent to which course meta information can improve model transferability, in addition to behavior features (RQ2), and whether fine-tuning the latter meta models on previous iterations of a course unseen during training leads to higher performance on the last iteration of the same course (RQ3). 
The dataset, optimization protocol, and experiments are described below.

\begin{table*}
\centering
\small
\resizebox{\textwidth}{!}{
\begin{tabular}{rrrrrrrrrrrrrrr} 
\toprule
\multirow{2}{*}{\textbf{Title}} & \multirow{2}{*}{\textbf{Identifier}} & \multicolumn{2}{l}{\textbf{Iterations}$^1$} & \multirow{2}{*}{\textbf{Topic}$^2$} & \multirow{2}{*}{\textbf{Level}$^{3^8}$} & \multirow{2}{*}{\textbf{Language}$^4$} & \multirow{2}{*}{\textbf{No. Weeks$^5$}} & \multicolumn{2}{l}{\textbf{No. Students}$^3$} & \multicolumn{2}{l}{\textbf{Passing Rate}$^6$ [\%]} & \multirow{2}{*}{\textbf{No. Quizzes}$^7$} \\
&  & \textit{Trn} & \textit{Trs} & & & & & \textit{Trn} & \textit{Trs} &\textit{Trn} & \textit{Trs} & &\\
\midrule
Comprendre les Microcontrôleurs & Micro & 4 & 0 & Eng & BSc & French & 10 & 3,974 & - & 26.9 & - & 18 \\ 
\hline
Analyse Numérique & AnNum & 3 & 0 & Math & BSc & French & 9 & 1,471 & - & 51.5 & - & 36 \\ 
\hline
Household Water Treatment and Storage & HWTS & 2 & 0 & NS & BSc & French & 5 & 2,438 & - & 47.2 & - & 10 \\ 
\hline
Programmation Orientée Objet & OOP & 1 & 0 & CS & Prop & French & 10 & 797 & - & 38.1 & - & 10 \\ 
\hline
Programmation en C++ & InitProgC++ & 1 & 0 & CS & Prop & English & 8 & 728 & - & 63.3 & - & 13\\ 
\hline
\hline
Digital Signal Processing & DSP & 4 & 1 & CS & MSc & English & 10 & 11,483 & 4,012 & 22.6 & 23.1 & 38\\ 
\hline
Villes Africaines & Villes Africaines & 2 & 1 & SS & BSc/Prop$^9$ & En/Fr$^9$ & 12  & 7,888 & 5,643 & 6.3 & 9.9 & 18 \\ 
\hline
L'Art des Structures I & Structures & 2 & 1 & Arch & BSc & French & 10  & 278 & 95 & 57.7 & 66.3 & 6\\ 
\hline
Functional Programming & ProgFun & 1 & 1 & CS & BSc & French & 7 & 11,151 & 7,880 & 50.72 & 81.33 & 3\\ 
\hline
\hline
Launching New Ventures & Venture & 0 & 1 & Bus & BSc & English & 7 & - & 6,673 & - & 1.4 & 13\\ 
\hline
Éléments de Géomatique & Geomatique & 0 & 1 & Math & BSc & French & 11 & - & 452 & - & 45.1 & 27\\
\bottomrule
\multicolumn{14}{l}{$^1$ \textbf{Set abbrev.} \textit{Trn}: training; \textit{Trs}: transfer.}\tabularnewline  
\multicolumn{14}{l}{$^2$ \textbf{Topic abbrev.}
\textit{Eng}: Engineering;
\textit{Math}: Mathematics;
\textit{NS}: Natural Science;
\textit{CS}: Computer Science; \textit{SS}: Social Science; \textit{Arch}: Architecture; \textit{Bus}: Economics and Business.} \tabularnewline
\multicolumn{14}{l}{$^3$ The values are computed after removing early-dropout students.$^4$\textbf{Level} is chosen by majority label in \textit{Trs} or \textit{Trn}. $^5$\textbf{Language} is chosen by majority label in \textit{Trs} or \textit{Trn}.} \tabularnewline
\multicolumn{14}{l}{$^6$ \textbf{Passing Rate} is averaged over the courses in \textit{Trs} or \textit{Trn} weighted by number of students. $^7$\textbf{No. Quizzes} is averaged over the courses in \textit{Trs} or \textit{Trn}. } \tabularnewline
\multicolumn{14}{l}{$^8$ \textbf{Level abbrev.} \textit{Prop}: Propedeutic / Other; \textit{BSc}: Bachelor; \textit{MSc}: Master. $^9$ For \textit{Villes Africaines}, the / operator represents characteristics of courses in \textit{Trn} / \textit{Trs}.}\tabularnewline 
\end{tabular}}
\vspace{0.4mm}
\caption{Detailed information about the courses included in our dataset.}
\label{tab:transfer}
\end{table*}

\begin{table*}[]
\small
\centering
\vspace{-4mm}
\begin{tabular}{c|cccc||cccc}
\toprule
 & \multicolumn{4}{c}{\textbf{40\% Early Prediction Level}} & \multicolumn{4}{c}{\textbf{60\% Early Prediction Level}} \\ 
 & \textcolor{gray}{\textbf{1-1 Same$^1$}} & \textbf{1-1 Diff$^2$} & \textbf{N-1 Same} & \textbf{N-1 Diff} & \textcolor{gray}{\textbf{1-1 Same$^1$}} & \textbf{1-1 Diff$^2$} & \textbf{N-1 Same} & \textbf{N-1 Diff}\\ \midrule
DSP & \textcolor{gray}{82.0} & {61.1} & {\textbf{83.1}} & 77.4 & \textcolor{gray}{92.7} & {65.3} & \textbf{91.8} & 87.8 \\ 
Villes Africaines & \textcolor{gray}{73.8} & {64.6} & {69.4} & \textbf{79.7} & \textcolor{gray}{82.9} & {67.0} & {80.7} & \textbf{82.7} \\ 
Structures & \textcolor{gray}{52.5} & {51.1} & {52.9} & \textbf{56.2} & \textcolor{gray}{55.2} & {51.3} & {50.4} & \textbf{54.4} \\ 
ProgFun & \textcolor{gray}{50.6} & {50.7} & \textbf{59.5} & {58.7} & \textcolor{gray}{50.8} & {51.0} & \textbf{62.3} & {62.0} \\
Ventures & \textcolor{gray}{51.0} & {{59.8}} & {-} & \textbf{70.3} & \textcolor{gray}{54.9} & {{60.2}} & {-} & \textbf{71.8} \\ 
Geomatique & \textcolor{gray}{76.2} & {57.7} & {-} & \textbf{68.4} & \textcolor{gray}{79.5} & {57.6} & {-} & \textbf{65.5} \\ \bottomrule
\multicolumn{9}{l}{\footnotesize For each course and early prediction level, the highest value on a transfer experiment is in \textbf{bold}.}\tabularnewline
\multicolumn{9}{l}{\footnotesize $^1$ \texttt{1-1 Same} is not included in determining the highest value across transfer courses, since it could not exist in a real-world setting.}\tabularnewline
\multicolumn{9}{l}{\footnotesize $^2$ The balanced accuracy is averaged across the five trained models.}\tabularnewline 
\end{tabular}
\vspace{1mm}
\caption{Behavior transfer performance in terms of balanced accuracy for each transfer course (the higher it is, the better).}
\label{tab:baseline}
\vspace{-4mm}
\end{table*}

\subsection{Dataset}
Our dataset includes $26$ MOOCs, taught by instructors of an European university between 2013 to 2015. The data is fully anonymized with regards to student information. In total, this dataset covers $145{,}714$ students. It contains fine-grained video and quiz interactions for each student, e.g., pressing pause on a video or submitting a quiz. After removing the early-dropout students (see Sec. \ref{sec:log-preproc}), our data set contains $73{,}042$ students in total. The variety of courses in the dataset (e.g., in terms of topic, duration, level, and language), the large scale, and the varied population (e.g., students come from different countries and cultures) allows us to provide a realistic estimation of model performance. The courses are spread across two languages, seven topics, and three university levels. The smallest course has $95$ students, whereas the largest course has $11{,}151$ students. Table \ref{tab:transfer} reports detailed information on each course as well as the assignment to train and transfer sets. We will use these assignments for \textbf{RQ1} and \textbf{RQ2}, i.e. we will use $20$ courses as the train courses in $C_{trn}$ ($48{,}287$ students in total), and $6$ courses as the transfer courses in $C_{trs}$ ($24{,}755$ students in total). To answer \textbf{RQ3}, we utilize another train-transfer split: we will hold out one course $C$ completely (i.e. all iterations of $C$ will be in $C_{trs}$) and train on all other courses. Courses are split in train and transfer sets to ensure diversity in topic, size, level, language, and passing rate.

\subsection{Optimization Protocol}
\label{sec:opt-proc}
We trained each model with features $H$ for the courses in $C_{trn}$ under two early prediction levels ($e=40\%$ and $e=60\%$), minimizing the binary cross-entropy loss according to the setting outlined for each research question in the next subsections. Grid search details are included in Appendix \ref{sec:opt-proc-2}. We monitored BAC due to the high class imbalance\footnote{We computed accuracy, F1 Score, AUC, precision, and recall in addition to BAC, but we found these metrics less representative of the predictive performance. 
}. Formally, we trained a model for each hyperparameter combination on the students in $S_{trn} = \cup \; S^c - \tilde{S}^c \; \forall \; c \in C_{trn}$ and then selected the best combination by measuring BAC on students in $S_{val}$. We then used the selected model to predict student success and compute BAC for each transfer course $c \in C_{trs}$.

\subsection{Behavior Transfer Evaluation (RQ1)}
\label{sec:rq1}
Our initial experiment is designed to investigate the extent to which student behavior transfers well on different iterations of the same course or of a different course. We considered four main settings:

\begin{itemize}[leftmargin=*,nolistsep]
\item \texttt{1-1 Same}. For each transfer course $c \in C_{trs}$, we optimized a \texttt{BO} model on a subset of the students ($80\%$) enrolled in the same course $c$. We then predicted on the remaining set of students ($20\%$) in the same course $c$, split into a test set and a validation set of $10\%$ each. We report results using a 10-Fold cross validation, stratified on the pass-fail binary outcome label. This setting is merely meant to give a measure of how difficult it is to predict on a course and is not applicable in the real world, given that we could not know in advance whether the students in the current iteration will pass or fail the course.
\item \texttt{N-1 Same}. For each transfer course $c \in C_{trs}$ (with $c \in C \subset \mathbb{C}$), we trained a \texttt{BO} model on the previous iterations of course $c$, i.e., $(C \; - \; \{c\}) \; \cap \; C_{trn}$. We then predicted on students of course $c$. 
\item \texttt{1-1 Diff}.  For each transfer course $c \in C_{trs}$ (with $c \in C \subset \mathbb{C}$), we optimized one separate \texttt{BO} model on each course $\tilde{c} \in C_{trs}$, with $c \neq \tilde{c}$. For each model, we then predicted on course $c$. 
\item \texttt{N-1 Diff}. For each transfer course $c \in C_{trs}$, we optimized a single \texttt{BO} model on all the training courses included in $C_{trn}$. We then predicted on students of course $c$. 
\end{itemize}

\vspace{2mm} \noindent The BAC achieved by our \texttt{BO} models on each combination of transfer course and training setting are listed in Table \ref{tab:baseline}. For the $40\%$ early prediction level (left section of Table \ref{tab:baseline}), we observe that transferring a model from one course to another (\texttt{1-1 Diff}) does not work well for any course. Training on previous course iterations (\texttt{N-1 Same}) works best for the \emph{ProgFun} and \emph{DSP} courses. We hypothesize that this observation is because the latter two courses include a large number of students ($> 15{,}000$) in their previous iterations and therefore, training the \texttt{BO} directly on the respective previous iterations leads to better predictive performance. Training a \texttt{BO} model on a large set of courses (\texttt{N-1 Diff}) on the other hand works best for \emph{Villes Africaines}, which is a course with a very low passing rate ($< 8\%$). We assume that in case of highly imbalanced courses, the \texttt{N-1 Diff} model leads to less biased predictions. The \texttt{N-1 Diff} model is also the best model for \emph{Structures}. This course contains a small number of students only (${\sim}100$ per iteration) and the passing rates for the different iterations vary a lot, which might explain the lower BAC of the \texttt{N-1 Same} model in comparison to the \texttt{N-1 Diff} model. We also observe that for \emph{Venture}, the \texttt{N-1 Diff} model by far outperforms a model trained on students from the course iteration (\texttt{1-1 Same}). \emph{Venture} is a course with a very low passing rate ($1.4\%$), likely leading to a high bias (towards failing students) in case of the \texttt{1-1 Same} model. The \texttt{N-1 Diff} model also performs well for \emph{Geomatique}, but we observe a transfer gap here with respect to a model trained on students of the same course iteration (\texttt{1-1 Same}).

For the $60\%$ early prediction level (right section of Table \ref{tab:baseline}), we make similar observations. The \texttt{N-1 Diff} represents the best setting for \emph{Villes Africaines} and \emph{Structures}. The \texttt{N-1 Diff} is again outperformed by the \texttt{N-1 Same} setting for courses with the highest number of students ($> 15{,}000$) and a reasonably high passing rate ($> 20\%$), i.e., \emph{ProgFun} and \emph{DSP}. Also at this early prediction level, the \texttt{BO} model trained under the \texttt{N-1 Diff} setting represents the best real world  solution on four out of six transfer courses.

We finally focus on the extent to which the balanced accuracy varies across courses under the best performing transfer settings. For courses with a very high number of students and a reasonably high passing rate, the best models trained under the \texttt{N-1 Same} setting show a high variance in performance ($40\%$ level: $83.1$ BAC for \emph{DSP} and $59.5$ BAC for \emph{ProgFun}; $60\%$ level: $91.8$ BAC for \emph{DSP} and $62.3$ BAC for \emph{ProgFun}). This difference in transfer performance could be justified by the fact that \emph{DSP} includes quizzes in the schedule (in contrast with \emph{ProgFun}, which only includes videos) and thus the behavior features are more discriminative for the course. The hypothesis that the lower performance is due to differences in the course design is also supported by the low BAC of the \texttt{1-1 Same} model, indicating that \emph{ProgFun} is a hard course to predict on. For courses with a low number of students and those with a high number of students but a low passing rate (\emph{Villes Africaines}, \emph{Structures}, \emph{Ventures}, and \emph{Geomatique}), the best models under \texttt{N-1 Diff} show a more stable transfer performance ($40\%$ level: avg. BAC $68.7$, std. dev. BAC $8.4$; $60\%$ level: avg. BAC $68.6$, std. dev. BAC $10.3$).

\begin{figure*}[t]
\vspace{-2mm}
\centering
\subfloat[40\% early prediction level \label{fig:noearly-40}]{
    \includegraphics[width=0.5\linewidth, trim=4 4 4 4,clip]{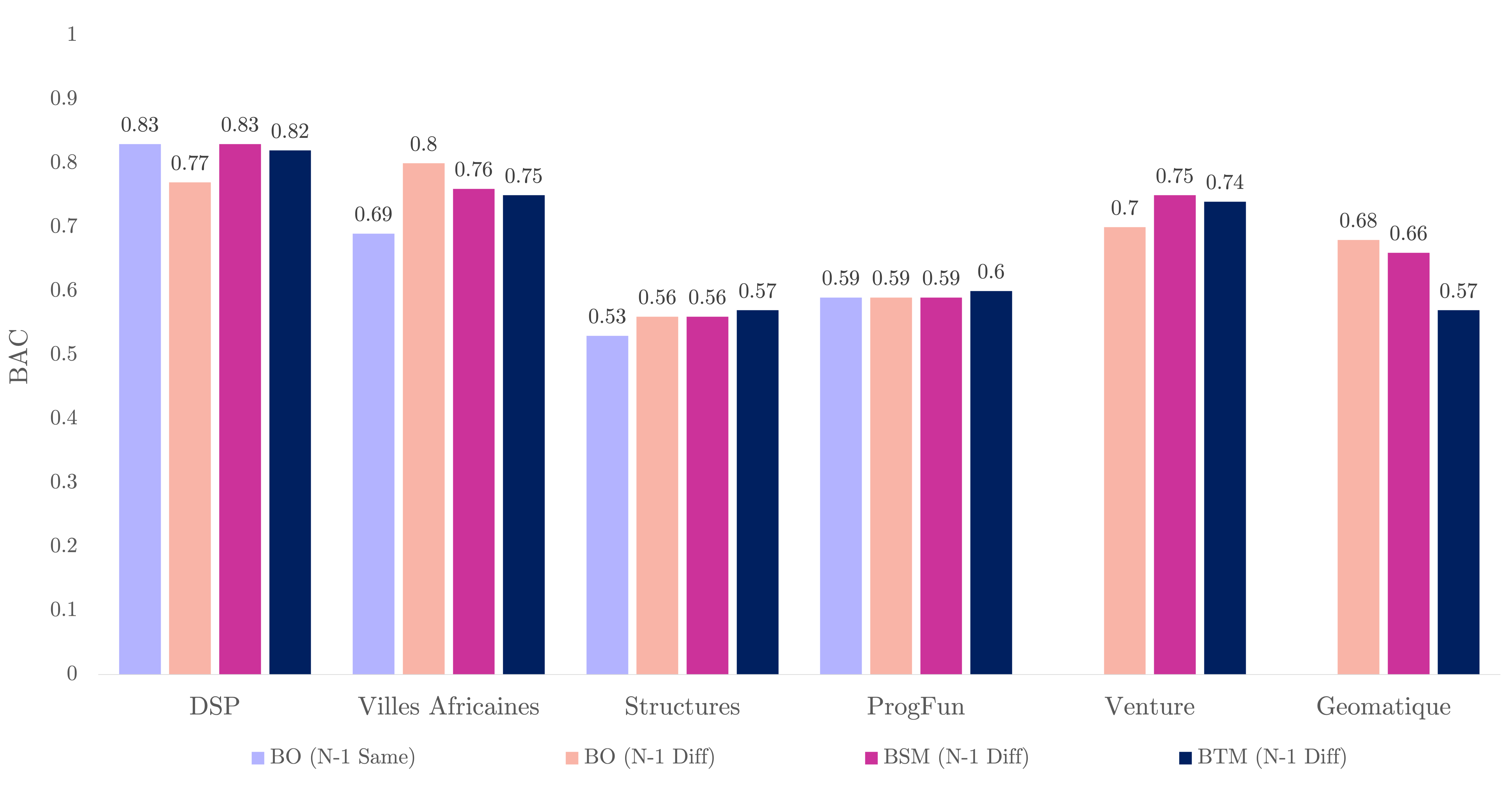}}
\subfloat[60\% early prediction level \label{fig:noearly-60}]{
  \includegraphics[width=0.5\linewidth, trim=4 4 4 4,clip]{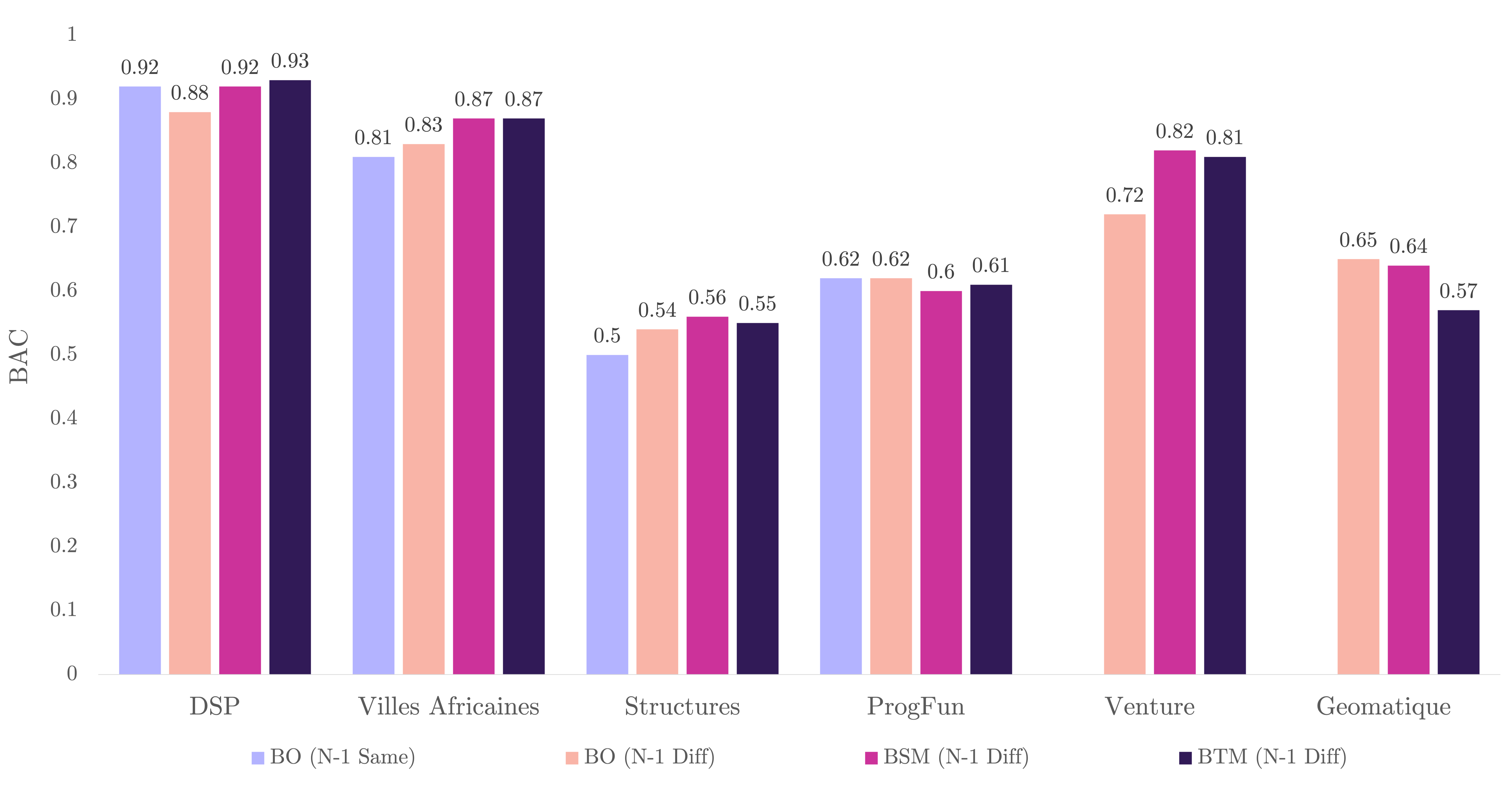}}
  \vspace{-3mm}
\caption{Comparison of the balanced accuracy achieved by our behavior only and behavior + meta models.}
\label{fig:meta-models}
\end{figure*}

In summary, for courses with previous iterations, a very high number of students, and a reasonably high passing rate, our \texttt{BO} model only on students from these previous iterations (\texttt{N-1 Same}) works the best. In the other cases, our \texttt{BO} model trained on students from the large set of courses (\texttt{N-1 Diff}) is the most accurate. On average, the behavior transfer BAC is $75\%$ (57\%) for courses with a reasonably high (low) number of quizzes, under both early prediction levels. On courses never seen during training, the behavior transfer BAC is around $70\%$ under both early prediction levels.

\subsection{Behavior + Meta Transfer Evaluation (RQ2)}\label{sec:results-metatransfer}
In a second experiment, we were interested in investigating the extent to which course meta information can improve model transferability. We therefore experimented with how to best utilize the meta information for our educational setting, trying different model architectures that pass in meta features at different locations of the model (see Section \ref{sec:model-creation}). We also identified which types of meta information were most relevant to the model through an ablation study. We considered four main settings for the courses, as follows:

\begin{itemize}[leftmargin=*,nolistsep]
\item \texttt{BO N-1 Same}. For each transfer course $c \in C_{trs}$ (with $c \in C \subset \mathbb{C}$), we optimized a \texttt{BO} model on the previous iterations of course $c$ included in the training set, i.e., $(C \; - \; \{c\}) \; \cap \; C_{trn}$. We then predicted on the students of course $c$.
\item \texttt{BO N-1 Diff}.  For each transfer course $c \in C_{trs}$, we optimized a single \texttt{BO} model on all the training courses included in $C_{trn}$. We then predicted on the students of course $c$. 
\item \texttt{BSM N-1 Diff}. For each transfer course $c \in C_{trs}$, we optimized a single \texttt{BSM} model on all the training courses included in $C_{trn}$. We then predicted on the students of course $c$. 
\item \texttt{BTM N-1 Diff}.  For each transfer course $c \in C_{trs}$, we optimized a single \texttt{BTM} model on all the training courses included in $C_{trn}$. We then predicted on the students of course $c$. 
\end{itemize}

\begin{figure*}[t]
\vspace{-7mm}
\centering
\subfloat[40\% early prediction level \label{fig:avg-40}]{
    \includegraphics[width=0.5\linewidth, trim=4 4 4 4,clip]{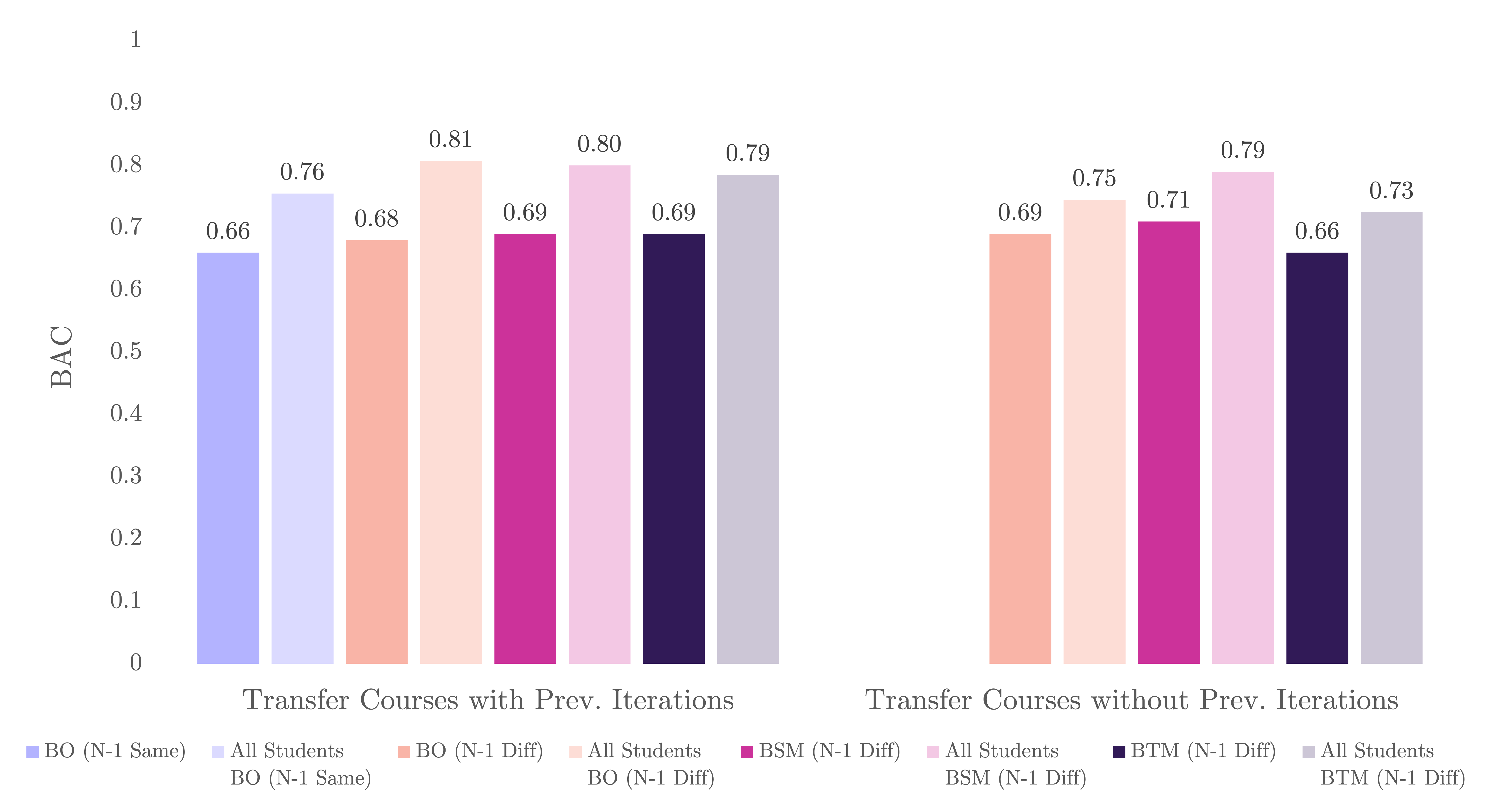}}
\subfloat[60\% early prediction level \label{fig:avg-60}]{
  \includegraphics[width=0.5\linewidth, trim=4 4 4 4,clip]{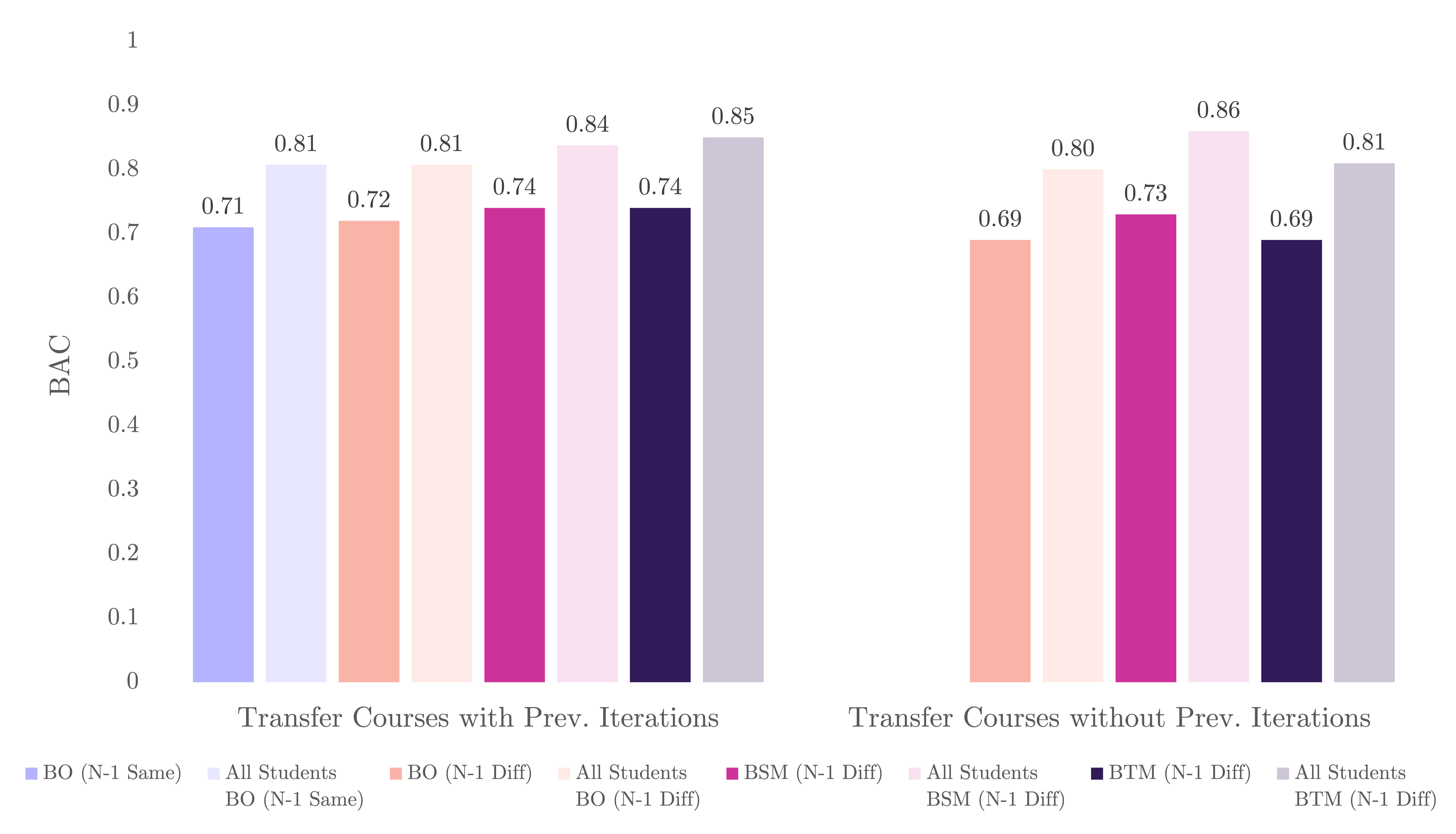}}
  \vspace{-2mm}
\caption{Average performance of \texttt{BO, BSM, BTM} across transfer courses with and without previous iterations, both on the filtered (w/ early-dropout students) and the entire student population.}
  \vspace{-3mm}
\label{fig:avg-meta-models}
\end{figure*}

\vspace{1mm} \noindent \textbf{Predictive Power Analysis}. Figure \ref{fig:avg-meta-models} illustrates the average predictive performance of our models across all transfer courses. We computed the average separately for courses with previous iterations in $C_{trn}$ and for courses that do not have previous iterations. We observe that adding meta information increases the average BAC for courses with previous iterations, no matter how that meta information is added: the \texttt{BSM (N-1 Diff)} and the \texttt{BTM (N-1 Diff)} both outperform the (\texttt{BO (N-1 Diff)}) model for both early prediction levels. For the courses without previous iterations we again observe the same pattern for the $40\%$ and $60\%$ early prediction levels. However, for these courses, the \texttt{BSM (N-1 Diff)} model performs better than the \texttt{BTM (N-1 Diff)} model. It therefore seems that for courses with previous iterations in $C_{trn}$, adding the maximal amount of meta features helps the model focus on the similar courses (i.e. the previous iterations of the transfer course) in the training data. In contrast, for models without previous iterations, it seems to be important to focus on only the relevant meta information (hence the better performance of the \texttt{BSM (N-1 Diff)} model. With respect to training on previous iterations, we observe that all the models trained on a large data set that include a range of courses (\texttt{BO (N-1 Diff)}, \texttt{BSM (N-1 Diff)}, \texttt{BTM (N-1 Diff)}) show the same or a higher BAC than the models trained on previous iterations of a course only (\texttt{BO (N-1 Same)}).

The shaded bars in Fig. \ref{fig:avg-meta-models} denote the predictive performance of our models for the whole student population (including early-dropout students). We observe that every one of our models achieve a higher BAC for all students in both early prediction settings. This result is expected as the early-dropout students tend show very low activity already at the beginning of the course, making it easy to predict on them. We also observe that the patterns of the average BAC are the same when predicting on the whole population versus on the filtered population. For the remaining experiments in this paper, we will therefore report the BAC of the different models only on the filtered population (as described in Sec. \ref{sec:log-preproc}).

Figure \ref{fig:meta-models} illustrates the BAC achieved by our models separately for each transfer course. For the $40\%$ early prediction level (Fig. \ref{fig:noearly-40}), it can be observed that adding meta information is helpful only for some courses of the courses with previous iterations. For \emph{DSP}, adding meta information leads to a strong increase in BAC: both meta models (\texttt{BSM N-1 Diff} and \texttt{BTM N-1 Diff}) outperform the model including behavior features only (\texttt{BO N-1 Diff}). For \emph{ProgFun} and \emph{Structures}, including meta information does not significantly increase the BAC (in the case of the \texttt{BSM N-1 Diff} and \texttt{BTM N-1 Diff} models). In the case of courses without previous iterations, adding meta information increases the BAC for one course (\emph{Ventures}), but leads to a slightly decreased predictive accuracy for the second course (\emph{Geomatique}). We further observe that the \texttt{BSM N-1 Diff} model tends to outperform the \texttt{BTM N-1 Diff} model. It performs better in four out of six courses. Furthermore, the \texttt{BSM N-1 Diff} model performs better than or at least as well as the model including behavior features only (\texttt{BO N-1 Diff}) for four out of the six courses. We observe a drop in performance only for two courses. Compared to the other transfer courses, \emph{Villes Africaines} has a high number of students and a highly imbalanced passing rate.
The second course, \emph{Geomatique}, seems to be hard to transfer to, as the performance of the \texttt{BSM N-1 Diff} model is very close to the performance of the \texttt{BO N-1 Diff} model. For courses with previous iterations, combining meta information and behavior features (\texttt{BSM N-1 Diff}) led to the same or a higher BAC as the model trained on previous iterations of the transfer course (\texttt{BO N-1 Same}).

For the $60\%$ early prediction level (Fig. \ref{fig:noearly-60}), the benefit of including meta information is more evident. Except for \emph{ProgFun} and \emph{Geomatique}, both meta models (\texttt{BSM N-1 Diff} and \texttt{BTM N-1 Diff}) have a BAC at least as high as the model trained on behavior features only (\texttt{BO N-1 Diff}. In contrast to the $40\%$ early prediction level, the performance differences between the \texttt{BSM N-1 Diff} model and the \texttt{BTM N-1 Diff} model are not as pronounced; the \texttt{BSM N-1 Diff} model outperforms the \texttt{BTM N-1 Diff} model for three courses, but performs (slightly) worse than the \texttt{BTM N-1 Diff} model for two other courses. Both the \texttt{BSM N-1 Diff} model and the \texttt{BTM N-1 Diff} model achieve or exceed the predictive performance of a model trained on previous course iterations (\texttt{BO N-1 Same}) for three out of four courses. For \emph{ProgFun}, the predictive performance of the two models is very close to that of the \texttt{BO N-1 Same} model.
     
Subsequently, we focus on the amount the BAC varies across courses under the best performing setting, namely \texttt{BSM N-1 Diff}. On average, the BAC is higher in all model settings for the $60\%$ early prediction level than for the $40\%$ level. This is expected because we would assume that the predictive power generally improves with more available data about the student behavior. We observe that the \texttt{BSM N-1 Diff} model shows a lower BAC for \emph{Structures} and \emph{ProgFun} in comparison to the other courses. We have already observed in Sec. \ref{sec:rq1} that it seems to be generally hard to predict on the selected iteration of the \emph{ProgFun} course (compare also the predictive performance of the \texttt{1-1 Same} model in Table \ref{tab:baseline}). Both \emph{Structures} and \emph{ProgFun} include a comparably low number of quizzes (see Table \ref{tab:transfer}), we therefore hypothesize that the presence of quizzes has a positive impact on model transfer.

\begin{figure*}[]
\vspace{-2mm}
\centering
\subfloat[$40\%$ early prediction level\label{fig:weights40-ablation}]{
   \includegraphics[width=0.5\linewidth, trim=4 4 4 4,clip]{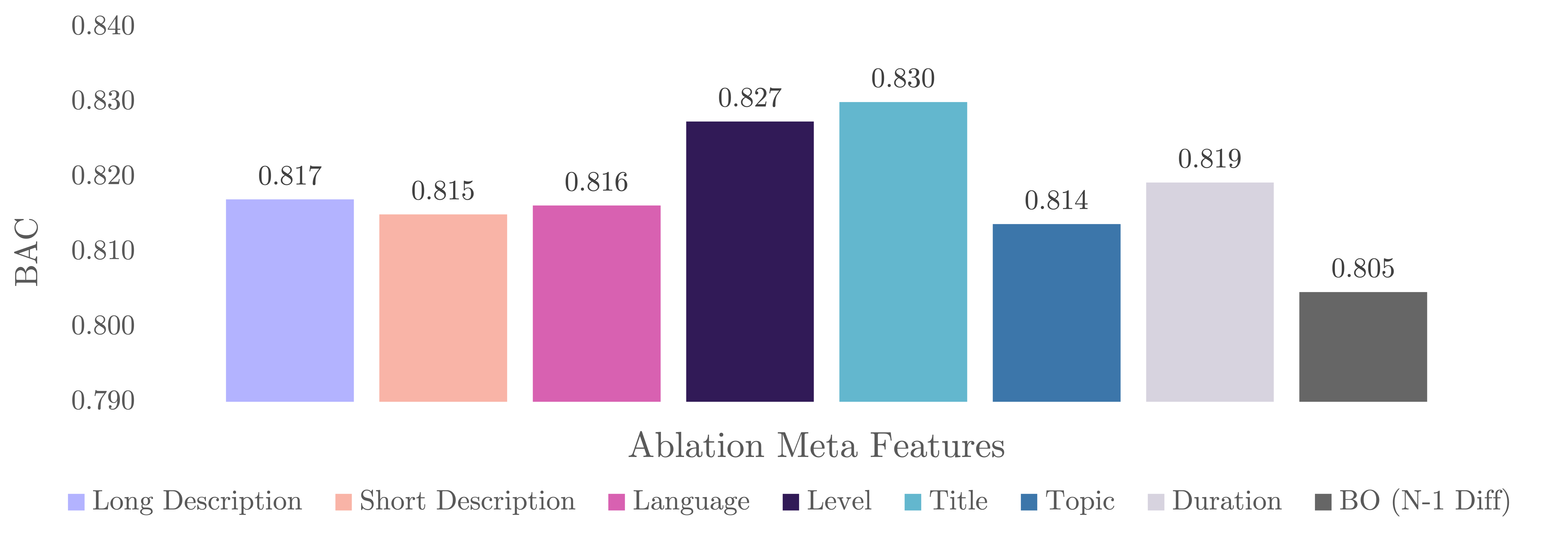}}
\subfloat[$60\%$ early prediction level\label{fig:weights60-ablation}]{
   \includegraphics[width=0.5\linewidth, trim=4 4 4 4,clip]{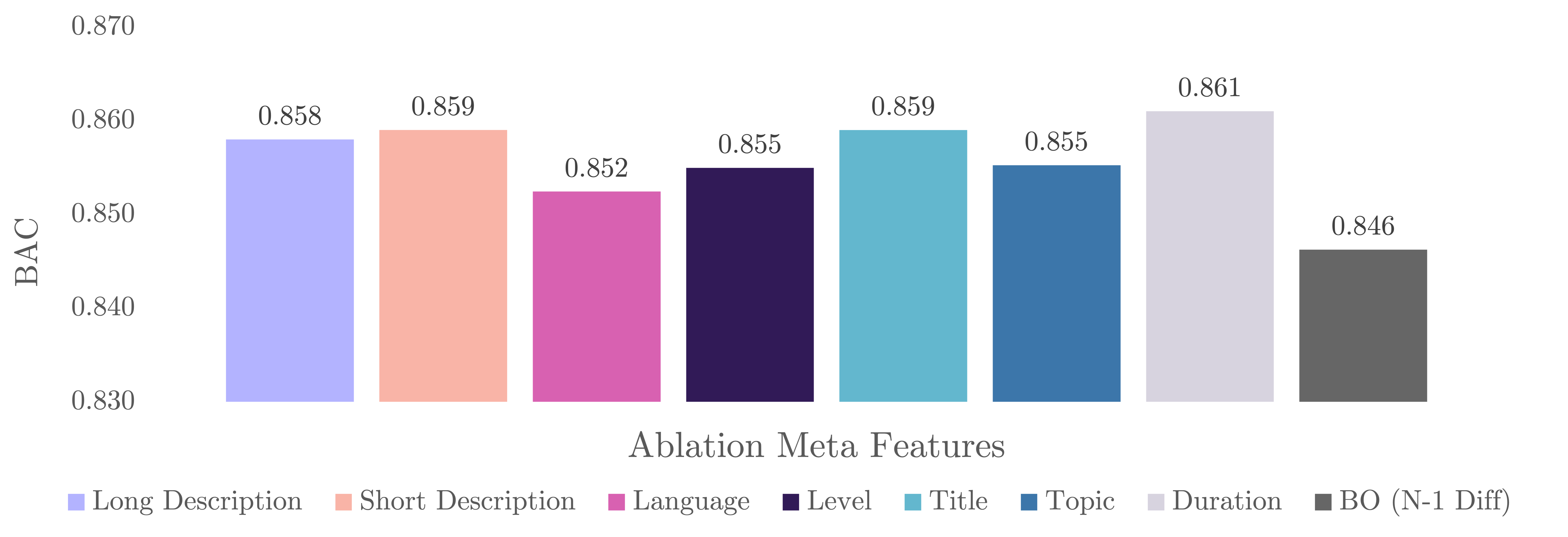}}
  \vspace{-3mm}
   \caption{Ablation study on the meta features for the  \texttt{BTM} model.}
\end{figure*}

\begin{figure*}
\centering
\includegraphics[width=1.0\linewidth]{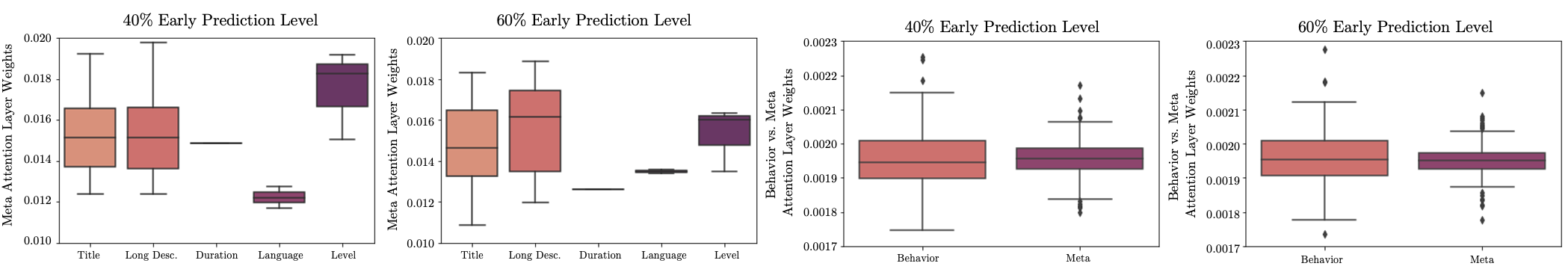}
  \vspace{-8mm}
\caption{Importance of the different meta features on the predictive power of the \texttt{BSM} model.}
\label{fig:weights}
\end{figure*}

\vspace{1mm} \noindent \textbf{Meta Information Importance Analysis}. To have a more detailed picture of the impact of the different meta features (see Section \ref{sec:features-descr}), we performed an ablation study. We trained a \texttt{BTM N-1 Diff} model on $80\%$ of the students attending the courses in $C_{trn}$ with only one meta feature at a time. We then computed the BAC of the trained model on the remaining $20\%$ of the students, split into a test and validation set. When performing the train-test split, we stratified by course and pass-fail label. Figures \ref{fig:weights40-ablation} and \ref{fig:weights60-ablation} illustrate the resulting BAC for the $40\%$ and $60\%$ early prediction levels. The rightmost bar represents the BAC of a \texttt{BO N-1 Diff} model (no meta information included). Notably, all the meta features improve the performance of the model over the baseline \texttt{BO N-1 Diff} on both early prediction levels. For the $40\%$ early prediction level, the most important meta features are represented by Title (with an embedding size of $60$), Level, and Duration. It therefore seems that Title contains enough information to help the model leverage patterns from courses about similar topics. Furthermore, the ablation importance of Level indicates that the interaction behavior of students might differ according to their school level (e.g. Bachelor vs. Master). We theorize that Duration is important since longer courses include more weeks of student behavior data at the $40\%$ early prediction level.
In the $60\%$ early prediction level, the most important meta feature is Duration: there is a minimal difference in importance for the other meta features. Based on the results of this ablation study, we include all meta features into our models. We however perform a grid search over Short/Long Description and Title (see Sec. \ref{sec:opt-proc}) to avoid training on highly redundant information.

To have additional insight into which meta features our model considers useful, we examine the two \emph{Attention} layers of our \texttt{BSM} model. By analyzing the importance weights resulting from these two layers, we can assess the impact of the different meta features in combination with the behavioral features on success prediction. We investigate these importance weights for the \texttt{BSM N-1 Diff} model trained on the courses in $C_{trn}$ with all the meta features. 
The importance weights are illustrated in Fig. \ref{fig:weights} for both layers and both the $40\% $and the $60\%$ early prediction levels. Considering the first \emph{Attention} layer for the $40\%$ early prediction level (two leftmost plots) which weights the meta features separately, it can be observed that on average, Level appears as the highest weighted meta features, followed by Title, Long Description, and Duration. As also observed in the ablation study, Language appears to have lower importance than the other meta features at this level. For the $60\%$ early prediction level, the weights assigned to Long Description and Language increase, whereas the weights of Duration and Level decrease. We hypothesize that when a model has less data about the students in the course ($40\%$ early prediction level), it relies on heuristics about Duration and Level for estimating the course structure and pass-fail patterns and making predictive decisions. When more information about the course is available ($60\%$ level), meta features like Long Description and Language become slightly more relevant as the similarity of the course to other courses in terms of these meta features is more useful to the model. Considering the second \emph{Attention} layer (two rightmost plots), weighting all meta features against all behavior features, it can be observed that the importance of the behavior and meta features is, on average, equal. However, under the $60\%$ early prediction level, the variance of the importance for both types of features is lower, i.e. there are less polarized cases where either behavior features only or meta features only are important for a specific student.  

\vspace{1mm} \noindent In summary, combining behavior with meta features in a student prediction model leads to consistently higher transfer performance across courses, regardless of the early prediction level. Even when previous iterations are available, the predictive power of meta models is comparable to that of a model trained on the previous iterations of that course.
In particular, a static combination (\texttt{BSM}) is more transferable than a time-wise combination (\texttt{BTM}). Level, Title, and Duration are important for very early predictions (before mid-course), while all the meta features have a similar importance, with Duration still being slightly more important, for later predictions.

\begin{figure*}[]
\centering
\vspace{-4mm}
\subfloat[$40\%$ early prediction level\label{fig:fine40-tuned}]{
   \includegraphics[width=0.5\linewidth, trim=4 4 4 4,clip]{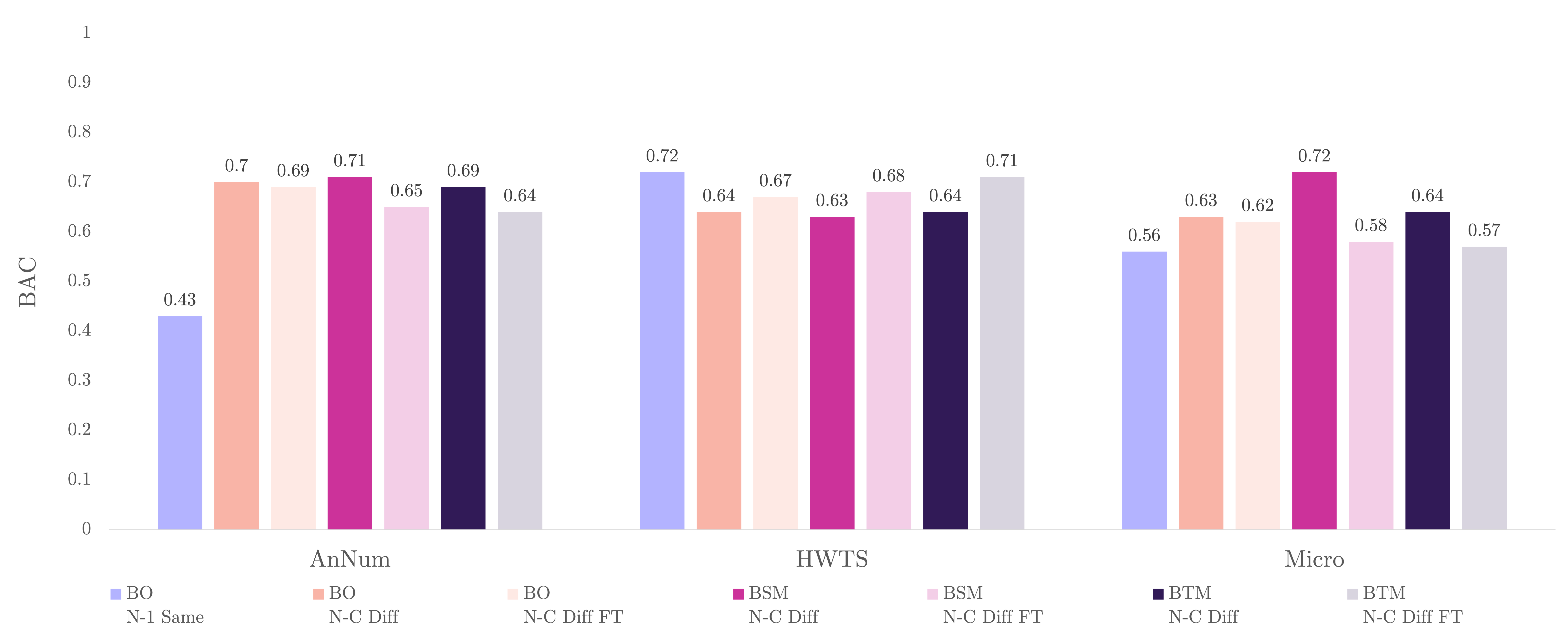}}
\subfloat[$60\%$ early prediction level\label{fig:fine60-tuned}]{
   \includegraphics[width=0.5\linewidth, trim=4 4 4 4,clip]{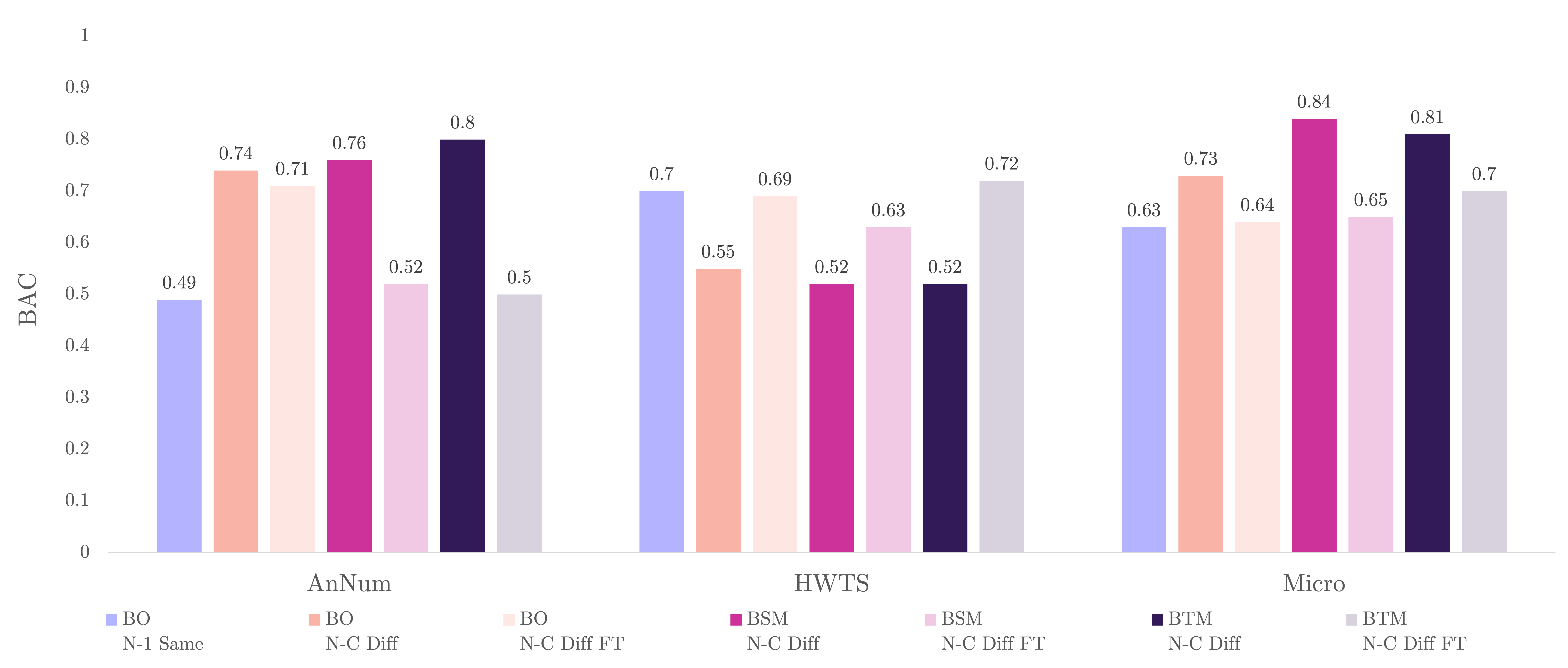}}
  \vspace{-3mm}
\caption{Comparative balanced accuracy of different fine-tuned meta models.}
\label{fig:fine-tuning}
\end{figure*}
\subsection{Transfer-by-Fine-Tuning Evaluation (RQ3)}
In a third and final experiment, we were interested in investigating the extent to which fine-tuning behavior and meta models on previous iterations of a course improves transfer. Related work from the broader computer vision community has demonstrated that fine-tuning can improve performance in transfer learning models \cite{guo2019spottune}. To form an analogous setup for this experiment, we held out a complete course $C$ (i.e. all iterations of that course) from the training data. In a second step, we fine-tuned the obtained model on a subset of iterations of $C$. Finally, we used this fine-tuned model to predict on the last iteration of $C$. We selected three courses with different characteristics (e.g., Duration, Topic) for our fine-tuning experiment and hence $C_{trs_{ft}} = \{Micro,  AnNum,  HWTS\}$.

\begin{itemize}[leftmargin=*,nolistsep]
\item \texttt{BO N-1 Same}. For each transfer course $c \in C_{trs_{ft}}$, we optimized a \texttt{BO} model on the previous iterations of course $c$, i.e. $C-c_{M^C}$.  We then predicted on the last iteration $c_{M^C}$ of course $C$. 
\item \texttt{BO N-C Diff}.  For each transfer course $C \in C_{trs_{ft}}$, we optimized a \texttt{BO} model on the training data set $(C_{trn}-C) \cup C_{trs}$. We then predicted on the students of the last iteration $c_{M^C}$ of course $C$. 
\item \texttt{BSM N-C Diff}. For each transfer course $C \in C_{trs_{ft}}$, we trained a  \texttt{BSM} model on the training data set $(C_{trn}-C) \cup C_{trs}$. We then predicted on the students of the last iteration $c_{M^C}$ of course $C$. 
\item \texttt{BTM N-C Diff}. For each transfer course $C \in C_{trs_{ft}}$, we optimized a \texttt{BTM} model on the training data set $(C_{trn}-C) \cup C_{trs}$. We then predicted on the students of the last iteration $c_{M^C}$ of course $C$.
\item \texttt{BO N-C Diff FT}. For each transfer course $C \in C_{trs_{ft}}$, we fine-tuned the optimal \texttt{BO N-C Diff} model (optimal on the training data set $(C_{trn}-C) \cup C_{trs}$) on a subset of iterations of that course $C-c_{M^C}$. We predicted on the last iteration $c_{M^C}$ of course $C$. 
\item \texttt{BSM N-C Diff FT}. For each transfer course $C \in C_{trs_{ft}}$, we fine-tuned the optimal \texttt{BSM N-C Diff} model (optimal on the training data set $(C_{trn}-C) \cup C_{trs}$) on a subset of iterations of that course $C-c_{M^C}$. We predicted on the last iteration $c_{M^C}$ of course $C$.
\item \texttt{BTM N-C Diff FT}. For each transfer course $C \in C_{trs_{ft}}$, we fine-tuned the optimal \texttt{BTM N-C Diff} model (optimal on the training data set $(C_{trn}-C) \cup C_{trs}$) on a subset of iterations of that course $C-c_{M^C}$. We predicted on the last iteration $c_{M^C}$ of course $C$.
\end{itemize}

\vspace{2mm} Figure \ref{fig:fine-tuning} illustrates the BAC of the different models for the three courses and two early prediction levels. For the $40\%$ early prediction level, we observe that fine-tuning is helpful for only one course. For \emph{HWTS}, fine-tuning on previous iterations of the course improves the BAC of behavior only (\texttt{BO N-C Diff}) and combined meta-behavior (\texttt{BSM N-C Diff} and (\texttt{BTM N-C Diff}) models significantly. For the other two courses, fine-tuning consistently decreases the predictive performance of all investigated model. For the $60\%$ early prediction level, the exact same patterns can be observed.

We postulate that the predictive performance increase achieved via fine-tuning depends on the similarity (of student behavior) between the different iterations of the course. The two iterations of \emph{HWTS} exhibit exactly the same course characteristics in terms of topic, language, duration, and level. Furthermore, both iterations of \emph{HWTS} contain a similar amount of students with a very similar (balanced) passing rate ($45.5\%$ vs. $45.9\%$). Therefore, training on previous iterations works well for this course; as demonstrated in Fig. \ref{fig:fine-tuning}, the \texttt{BO 1-Same} model outperforms the other models for both early prediction levels. For the other two courses, we observe that a model trained on previous iterations of the same course exhibits a suboptimal performance. In particular, the \emph{AnNum} course shows dissimilarities in terms of passing rates over the years, i.e. going from a very low passing rate in the first iteration ($8.5\%$) to passing rates $>70\%$ for iterations $2$ and $3$. Indeed, we observe that training a model on previous iterations does not work well for \emph{AnNum} (see BAC of the \texttt{BO 1-Same} model in Fig. \ref{fig:fine-tuning}), independent of the early prediction level. Finally, \emph{Micro} changed over subsequent iterations: the first run of the course started out with a lower number of students ($571$ students), but was almost balanced in terms of pass-fail labels (pass rate: $49\%$). The subsequent iterations saw a steep increase in student count ($>4{,}000$ students), but also a large drop in the pass rate (${\sim}5\%$). Training a model on a highly imbalanced course can lead to a bias (towards failing students), while fine-tuning on previous iterations of this course deteriorates model performance.


\section{Conclusions and Future Work}
In this paper, we investigated the transferability of models trained on different combinations of diverse student behavior and course meta features for early student success prediction. In contrast to prior work \cite{boyer2015transfer}, our study aimed to exploit course datasets at scale to cover student prediction on the last iteration of a given course and on courses with no previous iterations available. To do so, we introduced three novel strategies: 1) pre-training a model on a large set of diverse MOOCs, 2) including meta features using two different architectures, one combining behavior and meta features statically, and another one combining both in a time-wise fashion, and 3) fine-tuning the aforementioned models on previous iterations of courses. We then explored the extent to which student behavior transfers well on different iterations of the same course or of a different course, the extent to which the inclusion of course meta information can improve model transferability, and finally whether fine-tuning on previous iterations of a course unseen during training leads to higher performance on the last iteration of the same course. This is the first comprehensive evaluation of model transferability for \textit{early} success prediction. The key implications are as follows:

\begin{enumerate}[leftmargin=*]
\item Behavior transfer on this task depends on the number of previous course iterations, the number of students in those iterations, and their passing ratio. If a course has past iterations with a high number of students and a reasonably high passing rate, behavior features from the past iterations of that course works best. Otherwise, behavior features from a large and diverse course set leads to higher transferability. Few courses have low BACs across all model architectures (transfer or individually trained), which reflects low underlying signal in the data.
\item Combining behavior and meta features using attention layers results in the best transfer performance across courses, regardless of the early prediction level. The resulting models and the models trained on the previous iterations of a course (if available) have comparable or lower predictive power, confirming findings of prior research on meta transfer learning in different domains \cite{sun2019meta, qureshi2017wind, winata2020meta}. Our pre-trained static meta model could hence be seamlessly applied to predict on new unseen courses.
\item Predictive performance seems to be influenced by the number of quizzes in a course. For both behavior only and meta models, the more quizzes the transfer course included, the higher the BAC was. This implies that instructors who consider utilizing student success predictors should carefully design their course to include a good mix of lecture material and assessments. 
\item While all the meta features have similar importance for later prediction, the course Level, Title, and Duration meta features are important for very early predictions. Therefore, instructors should precisely define this course meta information. Educational data scientists should include such meta features in their models to foster transferability.  
\item When previous course iterations are available and similar (in student population and course structure) to the current transfer course, fine-tuning a meta model on these iterations can better transfer. However, if there are significant differences between iterations, fine-tuning the model will lead to a decrease in BAC.
\end{enumerate}

Our results indicate that behavior and meta features could be successfully used to improve model transferability. The models trained in this paper can be used to warm-start performance predictions, aiming to improve student performance in ongoing courses through targeted individual and curriculum interventions \cite{perez2021can}. Nevertheless, our work has several limitations that can be addressed with future research. While we tested a variety of MOOCs with different topics and populations on two early prediction levels, it is notable the MOOC dataset is from a single university in Europe and collected five years ago, leading to slight model bias because of the underlying data. Further work is needed to account for different course modalities, such as flipped classrooms and blended courses, to see the extent to which transferability is modality-agnostic. Additionally, our study relies on hand-crafted behavior features extracted from the raw clickstreams. Latent behavior feature models directly acting on clickstreams (e.g., autoencoders) can be combined with our architectures, to provide an end-to-end model and avoid the need of manually extracting behavior features. 

\vspace{2mm} \noindent \textbf{Acknowledgements}. We kindly thank SERI and Professor Martin Jaggi for supporting this project.

\balance 
\bibliographystyle{ACM-Reference-Format}
\bibliography{base}

\appendix
\section*{Appendix}
\begin{table*}
\small
\centering
\resizebox{0.94\linewidth}{!}{
\begin{tabular}{@{}lll@{}}
\toprule
\textbf{Set} & \textbf{Feature} & \textbf{Description} \\ \midrule
\multirow{3}{*}{\textit{Regularity}} & DelayLecture & The average delay in viewing video lectures after they are released to students. \\
& RegPeakTimeDayHour & The extent to which students' activities are centered around a particular hour of the day. \\
& RegPeriodicityDayHour & The extent to which the hourly pattern of user’s activities repeats over days. \\
 \midrule
\multirow{13}{*}{\textit{Engagement}} 
 & NumberOfSessions & The number of unique online sessions the student has participated in. \\  
 & RatioClicksWeekendDay & The ratio between the number of clicks in the weekend and the weekdays \\ 
 & AvgTimeSessions & The average of the student's time per session. \\
 & TotalTimeSessions & The sum of the student's time in sessions. \\ 
 & StdTimeSessions & The standard deviation of student's time in sessions. \\ 
 & StdTimeBetweenSessions & The standard deviation of the time between sessions of each user. \\
 & TotalClicks & The number of clicks that a student has made overall. \\
 & TotalClicksProblem & The number of clicks that a student has made on problems this week. \\
 & TotalClicksVideo & The number of clicks that a student has made on videos this week. \\
 & TotalClicksWeekday & The number of clicks that a student has made on the weekdays. \\ 
 & TotalClicksWeekend & The number of clicks that a student has made on the weekends. \\ 
 & TotalTimeProblem & The total (cumulative) time that a student has spent on problem events. \\ 
 & TotalTimeVideo & The total (cumulative) time that a student has spent on video events. \\ 
 \midrule
\multirow{22}{*}{\textit{Control}}
& TotalClicksVideoLoad & The number of times a student loaded a video. \\
& TotalClicksVideo & The number of times a student clicked on a video (load, pause, play, forward). \\
& AvgWatchedWeeklyProp & The ratio of videos watched over the number of videos available. \\ 
& StdWatchedWeeklyProp & The standard deviation of videos watched over the number of videos available. \\
& AvgReplayedWeeklyProp & The ratio of videos replayed over the number of videos available. \\
& StdReplayedWeeklyProp & The standard deviation of videos replayed over the number of videos available. \\ 
& AvgInterruptedWeeklyProp & The ratio of videos interrupted over the number of videos available. \\
& StdInterruptedWeeklyProp & The standard deviation of videos interrupted over the number of videos available. \\ 
& FrequencyEventVideo & The frequency between every Video action and the following action. \\
& FrequencyEventLoad & The frequency between every Video.Load action and the following action. \\
& FrequencyEventPlay & The frequency between every Video.Play action and the following action. \\
& FrequencyEventPause & The frequency between every Video.Pause action and the following action. \\
& FrequencyEventStop & The frequency between every Video.Stop action and the following action. \\
& FrequencyEventSeekBackward & The frequency between every Video.SeekBackward action and the following action. \\
& FrequencyEventSeekForward & The frequency between every Video.SeekForward action and the following action. \\
& FrequencyEventSpeedChange & The frequency between every Video.SpeedChange action and the following action. \\
& AvgSeekLength & The student's average seek length (seconds). \\
& StdSeekLength & The student's standard deviation for seek length (seconds). \\
& AvgPauseDuration & The student's average pause duration (seconds). \\
& StdPauseDuration & The student's standard deviation for pause duration (seconds). \\
& AvgTimeSpeedingUp & The student's average time using Video.SeekForward actions (seconds). \\
& StdTimeSpeedingUp & The student's standard deviation of time using Video.SeekForward actions (seconds). \\
 \midrule
\multirow{7}{*}{\textit{Participation}} 
& CompetencyStrength & The extent to which a student passes a quiz getting the maximum grade with few attempts. \\
& CompetencyAlignment & The number of problems this week that the student has passed. \\
& CompetencyAnticipation & The extent to which the student approaches a quiz provided in subsequent weeks. \\
 & ContentAlignment & The number of videos this week that have been watched by the student. \\ 
 & ContentAnticipation & The number of videos covered by the student from those that are in subsequent weeks. \\ 
 & StudentSpeed & The average time passed between two consecutive attempts for the same quiz. \\ 
 & StudentShape & The extent to which the student receives the maximum quiz grade on the first attempt. \\
 \bottomrule
\end{tabular}}
\vspace{1mm}
\caption{Descriptions of 45 behavioral features used in model training as discussed in Sec. \ref{sec:features-descr}.}
\label{tab:features}
\end{table*}

\section{Hyperparameter Tuning}
\label{sec:opt-proc-2}
This appendix section details the hyperparameter tuning decisions made in the model optimization protocol (Sec. \ref{sec:opt-proc}), provided for community replication. The batch size was set to $64$. We used the Adam optimizer, with an initial learning rate of $0.001$. For all models, the number of \emph{BiLSTM} layers and the number of units per \emph{BiLSTM} layer were optimized via grid search. For the \texttt{BSM} model, we extended the grid search to the number of \emph{Dense} layers and the number of hidden units per layer in the final cascade. For both the \texttt{BSM} model and the \texttt{BTM} models, the grid search was further extended to the meta features to be included in the model (all combinations of meta features) and to the size of the word embeddings for the Title, the Short Description, and the Long Description. Details about hyperparameter grids are reported in our code repository\footnote{\texttt{https://github.com/epfl-ml4ed/meta-transfer-learning}}. The selected hyperparameter combination is the one achieving the highest balanced accuracy (BAC) on a validation set composed of students in $S_{val} = \cup \; \tilde{S}^c \; \forall \; c \in C_{trn}$, with $\tilde{S}^c$ denoting $10\%$ randomly selected students for a given course $c$ (we stratified this selection on course and pass-fail label).

\section{Features}
\label{sec:features}
This appendix section will describe the behavioral features used in model training in more detail, as discussed in Sec. \ref{sec:features-descr}. Our four selected feature sets (\textit{Regularity}, \textit{Engagement}, \textit{Control}, and \textit{Participation}) are highlighted by previous work \cite{marras2021can} as strongly performant feature sets for MOOCs. In total, we have extracted the 45 features listed in Table \ref{tab:features}. The majority of the features come from the \textit{Control} and \textit{Engagement} feature sets. For the mathematical equations, we refer to the papers in which the features were initially introduced \cite{boroujeni2016quantify, lalle2020data, chen2020utilizing, marras2021can}. A detailed implementation of all features can be found in EPFL ML4ED's code repository\footnote{\texttt{https://github.com/epfl-ml4ed/flipped-classroom}}.

\end{document}